\documentstyle[floats,preprint,prd,aps,amsfonts]{revtex}
\tighten
\begin{document}
\begin{flushright}
\bf {GSI--94--40} \\
\bf {August 1995}
\end{flushright}
\vskip 0.2in
\begin{center}
\Large {A GLOBAL TEST OF}\\
\vskip 0.1in
\Large{QCD THEORIES FOR DIRECT PHOTON PRODUCTION}\\
\vskip 0.2in
\large {
 Engelbert Quack$^{\star}$,\\}
\vskip 0.1in
\small\it {Gesellschaft f\"{u}r Schwerionenforschung (GSI),
Postfach 110552, Darmstadt, D-64220\\
Germany}
\vskip 0.15in
\large{ Dinesh Kumar Srivastava$^{\ddag}$}
\vskip 0.1in
\small\it {Variable Energy Cyclotron Centre,
     1/AF Bidhan Nagar, Calcutta 700 064\\
     India\\
and\\
Gesellschaft f\"{u}r Schwerionenforschung (GSI),
Postfach 110552, Darmstadt, D-64220\\
Germany}
\end{center}

\renewcommand{\baselinestretch}{1.5} 
\parindent=20pt

\vskip 0.4in
\begin{center}
\bf {Abstract}\\
\end{center}

Direct photon production data at fixed target and collider experiments
from $pp$ and $\bar{p}p$ collisions are analysed in NLO QCD.
The results are grouped into three sets having $\sqrt{s}=20-30$ GeV,
630 GeV and 1.8 TeV. It is seen that at the lowest energies considered,
the theory systematically underpredicts the results for all the values of
the transverse momentum. At the intermediate energy of the UA2 experiment,
the data are more accurately described at high $p_T$, and at the
highest collider energies no significant underprediction is seen even at low
$p_T$, when using recent structure functions as measured at HERA.

\vskip 30pt

$^{\star}$ E-mail~ :~E.Quack@gsi.de

$^{\ddag}$ E-mail~ :~dks@vecdec.veccal.ernet.in, dinesh@veccal.ernet.in
\noindent
\vfill \eject

\newpage

  Experimental and theoretical studies of direct or prompt photon production
in hadron-hadron collisions at large transverse momenta have been pursued
extensively in the literature. This interest stems from the expectation that
direct photons may provide a clean test of perturbative QCD, and also give
valuable information about the parton distribution functions of the hadrons.
The production of direct photons in lowest order proceeds through the
annihilation ($q\overline{q}\,\rightarrow\,g\gamma$) and the Compton
($qg\,\rightarrow\,q\gamma$) processes. The Compton processes have the unique
distinction of being sensitive to the gluon structure function. At the same
time a fairly precise next to leading order QCD treatment is available for
the analysis \cite{aur}.

This offers us an opportunity to perform a global test of the QCD theories
for the direct photon production. Huston et al. \cite{hust} have recently
performed this task using CTEQ2M \cite{cteq} structure functions, and
reported that in general
the theory underpredicts the data. They find a general trend of a larger
deviation at lower $p_T$, and a smaller deviation at larger $p_T$ at all
the energies, which they claim should require an additional nonperturbative
$p_T$ broadening of the initial state parton distribution.

In this note, we will demonstrate how this apparent gap between theory and
data is strongly reduced when using more recent structure functions, and how
NLO QCD nicely describes the data in the region where this is to be expected.

This analysis relies on the NLO QCD calculation of direct photon production
by P.Aurenche et al.~\cite{aur}. This treatment was recently applied to the
data for direct
photon production from the Fermilab collider, see for instance \cite{vogel},
to the global analysis mentioned above \cite{hust}, as well as to an
overview of existing data in order to form a basis for the prediction of
direct photon production at future (LHC,RHIC) experiments \cite{cq}.

In a spirit similar to that of ref.\cite{aur,hust,vogel,cq}, we use the NLO
QCD theory but with a different parametrization of the structure functions,
we now compute the cross section of direct photon production in
$pp$ and $\bar{p}p$ reactions and compare our results with the existing data.
The available experimental results for direct photon production in
$pp$ and $\bar{p}p$ collisions can be broadly categorized into three groups,
depending on the centre of mass energy; $\sqrt{s}$ = 20-30 GeV, 630 GeV, and
1.8 TeV. This increase of energy corresponds to a decrease of the momentum
fraction $x$ which is probed by the photon production process inside the
hadron. The elementary parton--parton collision occurs at
$\hat{s} = x_1 x_2 s$, where $\sqrt{s}$ is the c.m.~energy of the
hadron--hadron reaction, and $x_1$ and $x_2$ are the momentum fractions of the
parton from the beam and target hadron, respectively. For a photon emitted
at $\theta = 90^o$ in the c.m.s., corresponding to a rapidity of $y$ = 0,
the expressions are symmetric in $x_1$ and $x_2$, which we then simply denote
by $x$ (kinematically, the process actually converts a quark into a photon
momentum). All of the data we consider here are taken around midrapidity.
Calculating the production process involves an integration over $x_1$, $x_2$
within the limits $x_{min} \leq x \leq 1$. For $y$ = 0, and using the
normalized transverse momentum variable $x_T = 2p_T/\sqrt{s}$, we have
\begin{equation}
 \label{eqxmin}
  x_{min} = \frac{x_T}{2-x_T} = \frac{p_T}{\sqrt{s}-p_T}
\end{equation}
Therefore, the data for the first group should be sensitive to parton
structure functions down to $x_{\rm min}$=0.16, whereas for the second group
this sensitivity should be  down to $x_{\rm min}$=0.02. For the highest energy
considered, the structure function is explored down to $x$=0.0067.

In our calculation, we use the structure functions $MRS D-'$ from
Martins, Stirling and Roberts \cite{mrs}, which provides a good fit to
recent HERA data. We first illustrate the difference to the $CTEQ2M$
structure function,
which was used previously in this context, by comparing both parametrizations
in the $x$ and $Q^2$ range which is relevant to the data we are considering.
Fig.~1 shows the parton distribution functions provided by $CTEQ2M$ and
$MRS D-'$ for values of $Q^2$ = (10 GeV)$^2$ and $Q^2$ = (100 GeV)$^2$,
which is about the range covered by the UA2 and Tevatron photon data. For the
gluon distribution, plotted in fig.~1a, one observes a strong effect of the
$Q^2$ evolution on both structure functions. However, the parametrizations
are also different. With decreasing $x$, which corresponds to going to smaller
values of $p_T$, more gluons are provided by $MRS D-'$ than by $CTEQ2M$.
Going to higher $\sqrt{s}$, corresponding to probing of smaller $x$,
similarly opens a
gap between the gluon luminosities provided by the two structure functions.
Since the low $p_T$ region is dominated by the Compton processes involving
gluonic fusion, the calculation which uses the $CTEQ2M$ structure function
therefore results in a lower cross section than the one using $MRS D-'$.
For a comparison, we also show the gluon density provided by the latest
(1995) parametrization $MRS G$. It has the same trend of providing more gluon
luminosity, even somewhat more pronounced than $MRS D-'$. For a better
comparison to existing calculations, we are going to use $MRS D-'$ in the
following calculation.
In fig.~1b we plot the valence ($u$ only) and the sea quark distributions.
Here, less difference is seen when comparing the two structure functions
$CTEQ2M$ and $MRS D-'$, which is of the order of the effect of the
$Q^2$ evolution. Correspondingly, the result for the photoproduction cross
section for both distributions shows only little difference. This applies to
the high $p_T$ region where $q\bar{q}$ annihilation is the dominating process.

We now turn to the results for direct photoproduction in $pp$ and $\bar{p}p$
reactions. For the three groups of energies, we show the comparison of the
NLO QCD calculation and the data in the form of the fractional difference
$(data-theory)/theory$ using the invariant cross section
\begin{equation}
E{d\sigma\over d^3p} = {1\over 2\pi p_t} {d\sigma\over dp_tdy} \; .
\end{equation}
The results are shown for the different energy regions with increasing energy
in fig.'s~1a--1c.

The low energy group, shown in fig.~1a, contains the data of NA24 \cite{na24},
WA70 \cite{wa70}, UA6 \cite{ua6} and E706 \cite{e706} as compared to our
result. Here and in the following, we extrapolated data taken in a finite
$y$ interval to $y=0$ by using the procedure of Aurenche and Whalley
\cite{blue}. This calculation as well as the following ones are carried out
at a fixed scale of
$Q^2=p_t^2/4$ as is that of Huston et al.~\cite{hust}. As discussed for
instance in
\cite{cq}, the dependence on the scale is weak in this domain and mainly
affects the overall normalization, rather than the slope. We also note at this
point that a possible overall uncertainty in the normalization of the data
will also not affect the slope, as is discussed in \cite{hust}.

What we find in the comparison with the low energy data is a general trend
of an underestimate of the data by the NLO calculation, however no
significant $p_T$ dependence of the discrepancy.

At intermediate energy, only the data of UA2 \cite{ua2} exist and are shown
in the comparison to theory in fig.~2b. As is seen in the plot, the data are
already better described by theory, and the trend of a better agreement at
high $p_T$ might be read off. However, the absolute deviation is reduced as
compared to the calculation based on the $CTEQ2M$ distributions
(apparently, the first point in fig.~2b is not contained in fig.~4 of
\cite{hust} or is outside of the scale).

In the last figure, 2c, we show our results compared to the collider data from
D0 \cite{d0} and CDF \cite{cdf}. Here, the agreement is satisfactory, and no
significant trend of a discrepancy increasing with decreasing $p_T$ can be
read off. The apparent discrepancy between our result and the corresponding
one of Huston et al.~has its origin in the lower gluon luminosity
provided by the $CTEQ2M$ parametrization in the $x$ region relevant for the
low $p_T$ data, as we
discussed before. It is also felt that the introduction of an additional
broadening of the initial state parton distribution by some value
$<\Delta k_T> \sim$ 1 GeV of presumably nonperturbative origin would not
have provided a resolution of this discrepancy. For the UA2 as well as for
the high energy data, $<\Delta k_T> \ll p_T$ at which the data are taken and
thus the effect of the additional $<\Delta k_T>$ broadening would be hardly
visible. For the data in the low energy region, on the other hand, the
underestimate of the data is present for all values of $p_T$.

We conclude that by using the recent structure function $MRS D-'$ (and even
more so by the latest parametrization $MRS G$), we find
that deviations are in general much less as compared to the results of
Ref.\cite{hust}, and follow the logical sequence of being large at lower
energies and smaller $p_T$ and becoming insignificant at the highest energy
over the entire range of $p_T$, for which the data are available.
This conclusion is corroborated by the differences in the structure functions
used at lower $x$ values. The behavior we find is precisely what is expected
from any NLO calculation in general, here ${\cal O}(\alpha \alpha_s^2)$.
At small $\sqrt{s}$, the NLO QCD calculation necessarily underpredicts the
data, then starts to give a quantitative prediction at sufficiently
high $\sqrt{s}$ first at high values of $p_T$, and finally provides a
satisfactory quantitative description of direct
photon production over the entire range of $p_T$ range at high $\sqrt{s}$.
\vspace*{1cm}

{\em Acknowledgments} \\
We gratefully acknowledge P.Aurenche and his group for providing us their
program, and to him and R.Baier for helpful discussions.
We thank the hospitality of the ECT* in Trento, where this work was
initiated, and of GSI, Darmstadt, where it was completed.

\newpage
\section*{Figure Captions}
\begin{description}
\item {1.)} Comparison of the two structure functions used in the
QCD analysis, MRS D-' and CTEQ2M, in the $x$ and $Q^2$ range relevant to the
collider experiment, $Q^2 \sim (p_T^{\gamma})^2$. Also shown is the most
recent parametrization MRS G. Fig.~1a: Density distribution of gluons,
fig.~1b: Valence and sea quark distribution.
\item {2a.)} Fractional difference between the NLO QCD calculation and data
taken at low energies, $\sqrt{s} \leq 30.6$ GeV.
\item {2b.)} Fractional difference between the NLO QCD calculation and data
taken at the intermediate energy of $\sqrt{s}$ = 630 GeV by UA2.
\item {2c.)} Fractional difference between the NLO QCD calculation and data
taken at the Tevatron at $\sqrt{s}$ = 1.8 TeV.
\end{description}
\clearpage


\newpage
\begin{thebibliography}{39}
%
\bibitem{aur}P. Aurenche, R. Baier, M. Fontannaz and D. Schiff,
Nucl. Phys. {\bf B286}, 509 (1987); {\bf B297}, 661 (1988);
P. Aurenche, R. Baier and M. Fontannaz, Phys. Rev. {\bf D42}, 1440 (1990)
%
\bibitem{hust} J.~Huston, E.~Kovacs, S.~Kuhlmann, H.~Lai, J.~Owens and
W.~Tung, Phys. Rev. {\bf D 51}, 6139 (1995)
%
\bibitem{cteq} J.~Botts et al., CTEQ Collab., Phys. Lett. {\bf B304}, 159
(1993)
%
\bibitem{vogel} M.~Gl\"uck, L.~Gordon, E.~Reya and W.~Vogelsang,
Phys. Rev. Lett. {\bf 73}, 388 (1994)
%
\bibitem{cq} J.Cleymans, E.Quack, K.Redlich and D.Srivastava, Preprint
GSI--94--55, to appear in Int. J. Mod. Phys. A, 1995
%
\bibitem{mrs}A.~D.~Martin, W.~J.~Stirling and R.~G.~Roberts,
Phys. Lett. {\bf 306B}, 145 (1993);
H.~Plothow-Besch, Comp. Phys. Comm. {\bf 75}, 396 (1993)
%
\bibitem{na24} C. De Marzo et al. (NA24 collaboration), Phys. Rev.
{\bf D36}, 8 (1987)
%
\bibitem{wa70}  M.~Bonesini et al. (WA70 collaboration) Z. f. Phys. {\bf
C38}, 371 (1988)
%
\bibitem{ua6} G.~Sozzi et al. (UA6 collaboration) Phys. Lett. {\bf
B317}, 243 (1993)
%
\bibitem{e706} G. Alverson et al. (E706 collaboration), Phys. Rev.
{\bf D48}, 5 (1993)
%
\bibitem{blue} P. Aurenche and M.~R.~Whalley, Rutherford Appleton
Laboratory preprint RAL-89-106 (unpublished)
%
\bibitem{ua2} R.~Ansari et al. (UA2 collaboration), Z. f. Phys.
{\bf C41}, 395 (1988); Phys. Lett. {\bf B 263}, 544  (1991)
%
\bibitem{d0} A.~Smith (D0 collaboration), private communication.
%
\bibitem{cdf} F.~Abe et al. (CDF collaboration), Phys. Rev. Lett. {\bf 73},
2662 (1994); erratum: Phys. Rev. Lett. {\bf 74}, 1891 (1995)
%

\end{thebibliography}
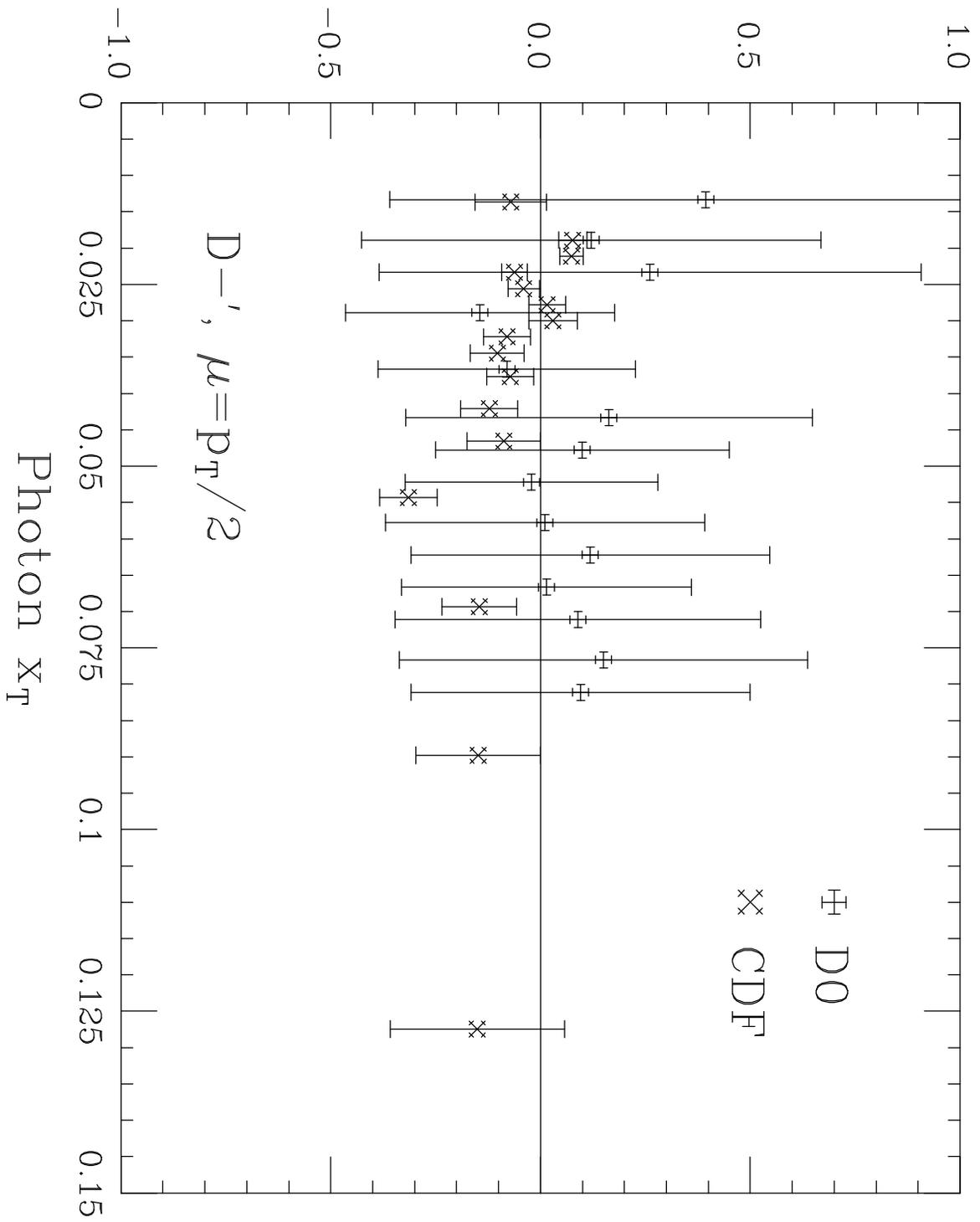
\end{document}